\documentclass[prd,superscriptaddress,amsfonts,amssymb,amsmath,showpacs,onecolumn]{revtex4}
\usepackage{bm}
\usepackage{amsfonts}
\usepackage{latexsym}
\usepackage[latin1]{inputenc}
\usepackage{graphicx}
\usepackage{amsmath}
\usepackage{palatino}
\usepackage{mathpazo}
\usepackage{textcomp}
\usepackage{float}
\usepackage{booktabs}
\usepackage{dcolumn}
\usepackage{hyperref}
\hypersetup{colorlinks,citecolor=blue}
\usepackage{amsmath}
\usepackage{xcolor}
\usepackage{mathtools}

\def\jnl@style{\it}
\def\aaref@jnl#1{{\jnl@style#1}}

\def\aaref@jnl#1{{\jnl@style#1}}

\def\aj{\aaref@jnl{AJ}}                   
\def\apj{\aaref@jnl{ApJ}}                 
\def\apjl{\aaref@jnl{ApJ}}                
\def\apjs{\aaref@jnl{ApJS}}               
\def\apss{\aaref@jnl{Ap\&SS}}             
\def\aap{\aaref@jnl{A\&A}}                
\def\aapr{\aaref@jnl{A\&A~Rev.}}          
\def\aaps{\aaref@jnl{A\&AS}}              
\def\mnras{\aaref@jnl{Mon.~Not.~Roy.~Astron.~Soc.}}             
\def\prd{\aaref@jnl{Phys.~Rev.~D}}        
\def\prc{\aaref@jnl{Phys.~Rev.~C}}  
\def\prl{\aaref@jnl{Phys.~Rev.~Lett.}}    
\def\qjras{\aaref@jnl{QJRAS}}             
\def\skytel{\aaref@jnl{S\&T}}             
\def\ssr{\aaref@jnl{Space~Sci.~Rev.}}     
\def\zap{\aaref@jnl{ZAp}}                 
\def\nat{\aaref@jnl{Nature}}              
\def\aplett{\aaref@jnl{Astrophys.~Lett.}} 
\def\apspr{\aaref@jnl{Astrophys.~Space~Phys.~Res.}} 
\def\physrep{\aaref@jnl{Phys.~Rep.}}      
\def\physscr{\aaref@jnl{Phys.~Scr}}       
\def\commat{\aaref@jnl{Comm.~Math.~Phys.}}              
\def\science{\aaref@jnl{Science}}               
\def\cqg{\aaref@jnl{Classical Quant.~Grav.}}            
\def\jpcs{\aaref@jnl{JPCS}}                                     
\def\ijmpd{\aaref@jnl{Int.~J.~Mod.~Phys.~D}}                    
\def\grg{\aaref@jnl{Gen.~Relat.~Gravit.}}               
\def\rpp{\aaref@jnl{Rep.~Prog.~Phys.}}          
\def\npa{\aaref@jnl{Nucl.~Phys.~A}}        
\def\lrr{\aaref@jnl{Living Rev.~Rel.}}                   
\def\jcap{\aaref@jnl{J.~Cosmology Astropart.~Phys.}}    
\def\rmp{\aaref@jnl{Rev.~Mod.~Phys.}}   

\allowdisplaybreaks[1]

\begin{document}
\color{black}       

\title{Constraining $f(R,T)$ Gravity From The Dark Energy Density Parameter $\Omega_{\Lambda}$}

\author{Snehasish Bhattacharjee }
\email{snehasish.bhattacharjee.666@gmail.com}
\affiliation{Department of Astronomy,
   Osmania University,
   Hyderabad-500007, India.}
\author{P.K. Sahoo}
\email{pksahoo@hyderabad.bits-pilani.ac.in}
\affiliation{Department of Mathematics, Birla Institute of Technology and
Science-Pilani, \\Hyderabad Campus, Hyderabad-500078, India.}
\begin{abstract}
$f(R,T)$ gravity is a widely used extended theory of gravity introduced in \cite{9} which is a straightforward generalization of $f(R)$ gravity. The action in this extended theory of gravity incorporates well motivated functional forms of the Ricci scalar $R$ and trace of energy momentum tensor $T$. The present manuscript aims at constraining the most widely used $f(R,T)$ gravity model of the form $f(R+2\lambda T)$ to understand its coherency and applicability in cosmology. We communicate here a novel method to find an lower bound on the model parameter $\lambda \gtrsim -1.9 \times 10^{-8}$ through the equation relating the cosmological constant ($\Lambda$) and the critical density of the universe ($\rho_{cr}$).

\end{abstract}

\maketitle


\section{Introduction}
The cosmological constant ($\Lambda$) problem is one of the major unsolved mysteries concerned with the dissimilarity between the tiny observed value of the cosmological constant and the extremely large value of zero point energy. Based on Planck energy cutoff along with other factors, the disaccord is as high as 120 orders of magnitude \cite{adler}, a predicament often quoted as \cite{2} "the worst theoretical prediction in the history of physics". After the discovery of the expansion of the universe by E. Hubble in 1929 \cite{hubble}, it was expected that the rate of expansion must be slowing down owing to the attractive nature of the gravity. Nonetheless, measurement of the intrinsic brightness of distant Type Ia supernovaes \cite{reiss, reiss2} showed that the expansion is in fact accelerating. This mysterious component which fuels the expansion at an ever increasing rate account for nearly 70\% of the energy budget of the universe and is termed Dark Energy (DE). There are three different DE models. These are: quintessence ($-1<\omega<0$), phantom energy ($\omega<-1$) and cosmological constant ($\omega=-1$), where $\omega$ represents EoS parameter. Current observations suggest $\omega\approx-1$ \cite{rebolo}.\\
Due to the lack of any observational evidence for the existence of any DE candidates (for a detailed reference of various DE candidates, one may refer to \cite{chaplygin}), researchers were inspired to modify the geometrical sector of the field equations where the Ricci scalar $R$ in the action is replaced by various generic functions of $f(R)$ \cite{7}, $f(\mathcal{T})$ \cite{8} where $\mathcal{T}$ represents Torsion scalar, $f(R,T)$ \cite{9} where $T$ is the trace of the energy momentum tensor, and $f(G)$ \cite{10} where $G$ is the Gauss-Bonnet invariant.  \\
Due to some fascinating features of $f(R,T)$ gravity and robustness is solving cosmological issues, it is often employed in the literature \cite{sahoo}. $f(R,T)$ gravity is also reported to clearly narrate the transition from matter dominated to late-time accelerated phase\cite{houndjo}. $f(R,T)$ gravity models have been applied to scalar field models \cite{scalar}, anisotropic models \cite{aniso1,aniso2}, dark matter \cite{in22}, dark energy \cite{in21}, bouncing cosmology \cite{bounce,bounce2}, gravitational waves \cite{in36}, super-Chandrasekhar white dwarfs \cite{in25}, massive pulsars \cite{in23}, wormholes \cite{in26,yousaf}, gravastars \cite{gravastar}, baryogenesis \cite{baryo,baryo2}, Big-Bang nucleosynthesis \cite{bang}, growth rate of matter fluctuations \cite{growth} and in varying speed of light scenarios \cite{physical}. In \cite{density}, the authors investigated the causes of irregular energy density in $f(R, T)$ gravity.\\
The present manuscript reports a pioneering method to constrain the model parameter of $f(R,T)$ gravity from the equation relating the cosmological constant ($\Lambda$) and the critical density of the universe ($\rho_{cr}$). According to the Friedmann solutions \cite{friedmann}, the critical density is a particular density at which the the universe is flat or Euclidean and as a result the curvature parameter vanishes \cite{peter}. The ratio of the current value of the density of the universe to the current value of the critical density is called the density parameter $\Omega_{0} = \rho/\rho_{cr}$. This is a very important cosmological parameter which determines the evolution and the ultimate fate of the universe \cite{peter}. For $\Omega_{0} = 1$, the universe is flat while for $\Omega_{0} > 1$ the universe is closed and open  for $\Omega_{0} <1$. Since current observations suggest $\Omega_{0} \simeq 1$, this indicates that the universe is approximately flat and apparently infinite and ergo favors the inflationary paradigm.\\
The paper is organized as follows: In Section \ref{II} we provide a summary of $f(R,T)$ gravity. In Section \ref{III} we introduce the framework to constrain the model parameter of $f(R,T)$ gravity. In Section \ref{IV} we present our conclusions.

\section{Overview of $f(R,T)$ Gravity}\label{II}

The action in $f(R,T)$ gravity is given by 
\begin{equation}\label{1}
S=\frac{1}{16\pi G} \int \sqrt{-g}\left[f(R,T) + \mathcal{L}_{m}\right] d^{4}x
\end{equation} 
where $\mathcal{L}_{m}$ denote matter Lagrangian.\\
Stress-energy-momentum tensor of matter fields reads 
\begin{equation}\label{2}
T_{\mu \nu} = \frac{-2}{\sqrt{-g}}\frac{\delta (\sqrt{-g} \mathcal{L}_{m} )}{\delta g^{\mu \nu}}
\end{equation}
varying the action (\ref{1}) with respect to the metric yields 
\begin{equation}\label{3}
\Pi_{\mu \nu}f^{1}_{,R}(R,T)+f^{1}_{,R}(R,T)R_{\mu \nu} -\frac{1}{2}g_{\mu\nu}f(R,T) =(T_{\mu \nu}+\Theta_{\mu \nu})+\kappa^{2}T_{\mu \nu}-f^{1}_{,T}(R,T)  
\end{equation}
where 
\begin{equation}\label{4}
\Pi_{\mu \nu}= g_{\mu \nu}\square-\nabla_{\mu} \nabla_{\nu}
\end{equation}
\begin{equation}\label{5}
\Theta_{\mu \nu}\equiv g^{\alpha \beta}\frac{\delta T_{\alpha \beta}}{\delta g^{\mu \nu}}
\end{equation}
and $f^{i}_{,X}\equiv \frac{d^{i}f}{d X^{i}}$. The field equations (\ref{3}) reduces to standard GR form when $f(R,T)\equiv R$.\\
Contracting equation (\ref{3}) with inverse metric $g^{\mu \nu}$, one obtain the trace of the field equations as 
\begin{equation}\label{6}
3\square f^{1}_{,R}(R,T)+f^{1}_{,R}(R,T)R -2f(R,T)= \kappa^{2}T-  (\Theta+T)f^{1}_{,T}(R,T). 
\end{equation}
Considering a spatially flat FLRW metric as \begin{equation}\label{7}
ds^{2}=dt^{2}-a(t)^{2}[dx^{2}+dy^{2}+dz^{2}]
\end{equation}
where $a(t)$ represents the scale factor. Assuming the universe to be a dominated by a perfect fluid and hence matter Lagrangian density can be assumed $\mathcal{L}_{m}=-p$. Applying this to equations (\ref{3}) and (\ref{6}) we obtain \begin{equation}\label{8}
\frac{\kappa^{2}+f^{1}_{,T}(R,T)}{f^{1}_{,R}(R,T)} \rho+\frac{1}{f^{1}_{,R}(R,T)}\left[pf^{1}_{,T}(R,T) -3 \dot{R} H  f^{2}_{,R}(R,T) + \frac{1}{2}\left( f(R,T)-Rf^{1}_{,R}(R,T)\right)  \right]=3H^{2}
\end{equation}
\begin{multline}\label{9} 
 \frac{\kappa^{2}+f^{1}_{,T}(R,T)}{f^{1}_{,R}(R,T)}p+ \frac{1}{f^{1}_{,R}(R,T)} \left[ \ddot{R} f^{2}_{,R}(R,T) +\dot{R}^{2} f^{3}_{,R}(R,T) -\frac{1}{2}\left( f(R,T)-R f^{1}_{,R}(R,T)\right)  -pf^{1}_{,T}(R,T) + 2H \dot{R}f^{1}_{,R}(R,T) \right] \\ =-3H^{2} -2\dot{H} 
\end{multline}
where dots represent time derivative and $H$ is the Hubble parameter, $\rho$ represents density and $p$ represents pressure with $T=\rho - 3p$.\\
We set the $f(R,T)$ functional form to be 
\begin{equation}\label{10}
f(R,T) = R + 2\lambda T.
\end{equation}
Substituting \eqref{10} in \eqref{8} we obtain the first modified Friedmann equation as 
\begin{equation}\label{11}
H^{2} = \frac{8 \pi G}{3} \left( 8 \pi + 3 \lambda\right) \rho - \frac{2}{3} \lambda \omega \rho
\end{equation}
where $\omega = p/\rho$ represents the EoS parameter. Current observations suggest $\omega\simeq-1$ \cite{rebolo}.
\section{Framework to Constrain $\lambda$}\label{III}
In this section we shall propose a framework to put bounds on the model parameter $\lambda$ from the equation relating the cosmological constant ($\Lambda$) and the critical density of the universe ($\rho_{cr}$).\\
We start by substituting $\omega=-1$ in \ref{11} to obtain 
\begin{equation}
\frac{(8 \pi)^{2} G}{3}\rho + \lambda \rho \left(8 \pi G + \frac{2}{3} \right)  = H^{2}.
\end{equation}
Since $8 \pi G\ll \frac{2}{3}$, we neglect the $8 \pi G$ term which simplifies the above equation to read
\begin{equation}\label{20}
 H^{2} =\frac{1}{3} \rho \left[ 2 \lambda + (8 \pi)^{2} G \right].
\end{equation}
Friedmann equation in standard GR with Cosmological Constant $\Lambda$ is given by \cite{sk,tp}
\begin{equation}\label{12}
H^{2} = \frac{8 \pi G}{3} \rho + \frac{\Lambda c^{2}}{3}.
\end{equation}
Since the L.H.S in Eqs. \ref{20} and \ref{12} are same, we can equate them to obtain 
\begin{equation}\label{13}
\Lambda c^{2} = 2 \lambda \rho + \left[(8 \pi)^{2} G - 8 \pi G \right] 
\end{equation}
Eq. \ref{13} can further be simplified to the reduced form 
\begin{equation}\label{14}
\Lambda \approx \frac{ \rho}{c^{2}}\left[2 \lambda + 192 \pi G \right] 
\end{equation}
Now the Cosmological Constant $\Lambda$ is defined as \cite{einstein}
\begin{equation}\label{15}
\Lambda= 3\left( \frac{H_{0}}{c}\right) ^{2} \Omega_{\Lambda}.
\end{equation}
Where $H_{0}$ is the present value of the Hubble parameter and $\Omega_{\Lambda}$ is the dark energy density parameter. As mentioned in the introduction that at the present epoch the total density parameter $\Omega_{0} = \rho/\rho_{cr} \simeq 1 $ \cite{planck}, we can therefore substitute $\rho$ with the critical density $\rho_{cr}$ in \ref{14}. This further yields 
\begin{equation} \label{16}
3\left( \frac{H_{0}}{c}\right) ^{2} \Omega_{\Lambda} \approx \frac{ 3 H_{0}^{2}}{8 \pi G c^{2}}\left[2 \lambda + 192 \pi G \right].
\end{equation}
Now the critical density $\rho_{cr}$ can be defined as \cite{friedmann}
\begin{equation}
\rho_{cr} = \frac{3 H_{0}^{2}}{8 \pi G}
\end{equation}
Simplifying \ref{16} we finally obtain 
\begin{equation}\label{18}
\Omega_{\Lambda} = \frac{1}{8 \pi G}\left( 2 \lambda + 192 \pi G\right) 
\end{equation}
\textbf{Observational Constraint:}
Planck satellite data reported $\Omega_{\Lambda} = 0.6889 \pm 0.0056$ \cite{planck}. This impose an lower bound on the model parameter $\lambda \gtrsim -1.9 \times 10^{-8}$.
\section{Conclusions}\label{IV}

$f(R,T)$ gravity is a modified theory of gravity where the Ricci scalar $R$ in the action is replaced by a generic function of $R$ and $T$ where $T$ denote the trace of energy momentum tensor. As a result, the emergent theory suffices major cosmological enigmas such as the current accelerated phase of the universe without requiring dark energy. $f(R,T)$ gravity is also reported to clearly narrate the transition from matter dominated to late-time accelerated phase\cite{houndjo}. \\
However the functional form of $f(R,T)$ can be arbitrary and any number of model parameters can be included which can be fine tuned to fit the observations. Hence we aimed in this paper to constrain the model parameter $\lambda$ for the simplest minimal coupled model $f(R,T) = R + 2 \lambda T$ through the equation relating the cosmological constant ($\Lambda$) and the critical density of the universe ($\rho_{cr}$). We report a lower bound on $\lambda \gtrsim -1.9 \times 10^{-8}$.\\
From the analysis we establish the idea that the parameter $\lambda$ is trivial and has no significant importance in cosmological models. It will be interesting to apply this method to constrain the model parameters of other $f(R,T)$ gravity models and to investigate their cosmological viability. 

\section*{Acknowledgments}

One of the author (PKS) acknowledge DST, New Delhi, India for providing facilities through DST-FIST lab, Department of Mathematics, BITS-Pilani, Hyderabad Campus where a part of the work was done. We are very much grateful to the honorable referee and the editor for the illuminating suggestions that have significantly improved our work in terms of research quality and presentation.

\end{document}